\documentclass[sigconf]{acmart}
\AtBeginDocument{%
  \providecommand\BibTeX{{%
    \normalfont B\kern-0.5em{\scshape i\kern-0.25em b}\kern-0.8em\TeX}}}

\copyrightyear{2024}
\acmYear{2024}
\setcopyright{acmlicensed}\acmConference[ACE 2024]{Australian Computing Education Conference}{January 29-February 2, 2024}{Sydney, NSW, Australia}
\acmBooktitle{Australian Computing Education Conference (ACE 2024), January 29-February 2, 2024, Sydney, NSW, Australia}
\acmPrice{15.00}
\acmDOI{10.1145/3636243.3636248}
\acmISBN{979-8-4007-1619-5/24/01}

\begin{document}

\title[Generative AI and Help-Seeking]{The Effects of Generative AI on Computing Students’ Help-Seeking Preferences}


\author{Irene Hou}
\affiliation{%
  \institution{Temple University}
  \city{Philadelphia, PA}
  \country{USA}}
\email{irene.hou@temple.edu}
\orcid{0009-0008-0511-7685}

\author{Sophia Mettille}
\affiliation{%
  \institution{Temple University}
  \city{Philadelphia, PA}
  \country{USA}}
\email{sophia.mettille@temple.edu}
\orcid{0009-0009-9707-6311}

\author{Zhuo Li}
\affiliation{%
  \institution{Temple University}
  \city{Philadelphia, PA}
  \country{USA}}
\email{tuq23149@temple.edu}
\orcid{0009-0000-5367-1076}

\author{Owen Man}
\affiliation{%
  \institution{Temple University}
  \city{Philadelphia, PA}
  \country{USA}}
\email{owen.man@temple.edu}
\orcid{0009-0003-0527-1395}

\author{Cynthia Zastudil}
\affiliation{%
  \institution{Temple University}
  \city{Philadelphia, PA}
  \country{USA}}
\email{cynthia.zastudil@temple.edu}
\orcid{0000-0002-3590-6975}

\author{Stephen	MacNeil}
\affiliation{
  \institution{Temple University}
  \city{Philadelphia}
  \state{PA}
  \country{United States}}
\email{stephen.macneil@temple.edu}
\orcid{0000-0003-2781-6619}

\newcommand{\td}[1]{{\color{black} #1}}
\newcommand{\fb}[1]{{\color{black} #1}}

\renewcommand{\shortauthors}{Hou, et al.}

\begin{abstract}
Help-seeking is a critical way that students learn new concepts, acquire new skills, and get unstuck when problem-solving in their computing courses. The recent proliferation of generative AI tools, such as ChatGPT, offers students a new source of help that is always available on-demand. However, it is unclear how this new resource compares to existing help-seeking resources along dimensions of perceived quality, latency, and trustworthiness. In this paper, we investigate the help-seeking preferences and experiences of computing students now that generative AI tools are available to them. We collected survey data (n=47) and conducted interviews (n=8) with computing students. Our results suggest that although these models are being rapidly adopted, they have not yet fully eclipsed traditional help resources. The help-seeking resources that students rely on continue to vary depending on the task and other factors. Finally, we observed preliminary evidence about how help-seeking with generative AI is a skill that needs to be developed, with disproportionate benefits for those who are better able to harness the capabilities of LLMs. We discuss potential implications for integrating generative AI into computing classrooms and the future of help-seeking in the era of generative AI. 
\end{abstract}







\maketitle

\section{Introduction}

Not so long ago, the internet ushered in a new wave of help-seeking resources for students~\cite{sharif2004undergraduates}. Students could ask questions and receive answers crowdsourced from experts around the world on Stack Overflow~\cite{dondio2020stackoverflow, nasehi2012makes, vasilescu2013stackoverflow} and find tutorials and lectures on YouTube~\cite{buzzetto2014youtube}. Over the last year, a new and widely available resource has emerged in the form of generative AI tools, such as GPT-3 and ChatGPT. A recent study revealed  GPT-3 has an ability to produce code explanations that students rated higher than those crafted by their peers~\cite{leinonen2023comparing}. This raises questions into the extent and circumstances under which students will turn to these tools for assistance. Especially as many other use cases are rapidly emerging in computing education contexts~\cite{macneil2022automatically, sarsa2022automatic,denny2023conversing, becker2023programming, leinonen2023using, savelka2023can, savelka2023thrilled, finnie2022robots, kazemitabaar2023studying, prather2023its, prather2023robots}. As a result, students now have instant access to high-quality and personalized help without needing to ask questions publicly to the entire class. 

It is critical to investigate the potential impact, positive and negative, these tools might have on the help-seeking preferences of students in computing education classes. We hope to gain a crucial preliminary understanding of when and why computing students choose to seek help and what resources they prefer. In particular, this paper aims to identify advantages and potential barriers to Generative AI as a help-seeking resource. This timely insight into how traditional help-seeking may have evolved with the emergence of generative AI will guide researchers in adapting their pedagogy to most effectively leverage this new help-seeking resource.

This paper explores computing students' preferences regarding generative AI. Through a mixed-methods study, we surveyed 47 students and conducted 8 interviews with additional students. Our goal was to understand the factors that influence help-seeking preferences and help resource utilization. Participants were asked to compare their experiences and preferences with traditional resources to their use of generative AI tools such as ChatGPT and GitHub Copilot. 
We investigated the following research questions:

\begin{enumerate}
    \item [\textbf{RQ 1:}] How does the frequency of help-seeking resource usage vary across tasks for computing students after the advent of Generative AI?
    \item [\textbf{RQ 2:}] What factor(s) may have influenced computing students’ decisions in using (or not using) generative AI tools over other existing resources?
    \item [\textbf{RQ 3:}] How do generative AI tools compare with other resources along perceived factors like trust, latency, and quality?
\end{enumerate}

Our research uncovered varied utilization patterns and preferences concerning generative AI tools. Students highlighted support for iteration, convenience, and the mitigation of social pressures as primary incentives for their adoption of these tools for help-seeking. However, the range of preferences and experiences seems closely tied to their ability to use these tools. Students who could assess generative model responses and integrate them into their problem-solving process appeared to derive the most value.

\section{Related Work} 

\subsection{Help-Seeking Behaviors and Challenges}

Effective help-seeking is crucial for academic success, as it often correlates with higher academic achievement and greater academic self-confidence~\cite{ryan2011help}. Help-seeking is a complex process and successful students tend to demonstrate proficient metacognitive and self-regulatory strategies~\cite{bergin2005examining, karabenick2013help}. However, help-seeking can be challenging for many. Students can encounter socio-emotional barriers to help-seeking~\cite{foong2017online} and decision-making barriers related to which help resources to utilize and when~\cite{cheng2011an, aleven2003help}.

These barriers guide whether students decide to seek help, which resources to consume, and when they engage with these resources such as peers, instructors, or online resources~\cite{foong2017online, wirtz2018resource, price2017isnap, karabenick2003seeking, newman1990childrens}.

When students seek help, they often encounter socio-emotional barriers. For example, the fear of appearing incompetent, worries about being burdensome, and the anxiety of potential rejection can significantly reduce their willingness to seek help~\cite{karabenick2003seeking, kumrow2007evidence}. 
Other fears stem from social capital; students may fear harming the trust and relationships they have cultivated with the peers in their community. While knowledge exchanged through social interactions can be invaluable, students may feel that knowledge comes at a social cost~\cite{chiu2006understanding}. 

Students also face decision-making barriers when choosing which help resources to utilize. Students must implicitly or explicitly evaluate the quality of the feedback source, determine receptivity of help providers, and formulate effective help requests~\cite{cheng2011an, aleven2003help}. Additionally, help-seeking is not a static decision-making process. Students tend to iteratively engage with the most convenient help resources first, then move on to scarcer information resources~\cite{wirtz2018resource, doebling2021patterns}. This iterative process has been called a ``recursive try-again'' loop~\cite{herring2016academic}. Students engage with increasingly effort-intensive resources until their need for help has been met. This iterative process is further supported by a study which found that students begin their help-seeking process with informal (i.e., detached) resources like peers or internet resources, to more formal (i.e., attached) resources like office hours or classrooms~\cite{wirtz2018resource}. This behavior suggests an underlying narrative that students are likely to prioritize help-seeking resources based on their current stage in the learning process and the resource's perceived quality and availability, while minimizing the barriers they face.

\subsection{Help-Seeking in Computing Education}

Undergraduate computing students are exposed to a wide variety of academic resources, both formal and informal; this includes office hours with TAs and instructors, peers online and offline, internet forums such as StackOverflow, course materials, and online resources~\cite{wirtz2018resource}.
When learning how to program there are a variety of challenges that students face which can be highly demoralizing. For example, they may struggle with exceedingly common syntactic issues, conceptual misunderstandings which impact their mental models of code and how it runs, and strategic knowledge about how to develop solutions to programming tasks, amongst other issues~\cite{qian2017students}. Another frustrating yet common sticking point for computing students is learning how to identify and resolve bugs~\cite{michaeli2019improving, fitzgerald2008debugging}. Navigating when and how to seek help during debugging presents a unique challenge for programming students. Moreover, the widespread availability of internet resources and prevailing industry values may play a role in how these student help-seeking behaviors differ compared to other fields. Distinctively, there is often a focus on troubleshooting and self-directed exploration, fueled by the abundance of digital resources and industry emphasis on self-directed learning~\cite{zander2012self, mccartney2016why}. 

Doebling and Kazerouni studied how computing students accessed help resources and found students start with online resources then move on to peers and finally to instructors only if necessary~\cite{doebling2021patterns}. This shows how students prioritize self-learning and rely on other sources as a next step, further corroborating the idea that students may feel the need to solve problems they encounter on their own before reaching out to other help sources. 
When using online resources, student tend to ask specific technical questions and issues that require quick solutions rather than more conceptual questions. Additionally, students view the use of online help resources more like a tool to utilize continuously throughout the coding process. More conceptual questions are reserved for when students are requesting help from sources such as their peers and instructors~\cite{doebling2021patterns}. The prioritization online tools by computing students when seeking help, coupled with the surging popularity of generative AI, could impact which resources students favor and when they choose to use them.

\subsection{Generative AI in Computing Education} 

Recently, there has been significant excitement about the potential for using generative AI in computing classrooms~\cite{prather2023transformed, lau2023from, zastudil2023generative} 
with researchers already demonstrating many remarkable capabilities of generative AI including generating code explanations~\cite{macneil2022experiences, macneil2022generating, leinonen2023comparing}, enhancing error messages~\cite{leinonen2023using}, identifying bugs~\cite{macneil2023decoding}, solving multiple-choice programming problems~\cite{savelka2023can, savelka2023thrilled}, solving Parsons problems~\cite{hou2023more}, and creating instructional materials and programming assignments~\cite{finnie2022robots, sarsa2022automatic, tran2023using}. 
A study on the impacts of chatbots on learning performance and motivation found students were motivated to receive help from a chatbot because it supported self-paced learning~\cite{yin2021conversation}.
Recently, a study showed that LLM-generated explanations of code even surpassed that of peers in a computing classroom~\cite{leinonen2023comparing}. Another study investigating how generative AI tools respond to student help requests have noted that GPT-3.5 in particular is helpful in identifying problematic areas of student code~\cite{hellas2023exploring}. LLMs have also been shown to be effective at providing on-demand, scalable programming assistance in computing education classes~\cite{liffiton2023codehelp}. \td{Previous studies have also identified that novice programming students write better code when supported by LLMs without decreasing their ability to manually modify code~\cite{kazemitabaar2023studying}. This is promising for help-seekers as they can solve problems that might otherwise be outside of their ability without losing their agency to adapt and revise the code themselves. }

Given these capabilities ranging from personalized support to timely feedback, generative AI tools could emerge as a robust source of help for computing students. However, whether or not students prefer these AI tools in comparison to other resources is a question that has been left largely unexplored.
Research is needed to understand how and why students rely on these tools in the context of other existing help resources. 

\section{Methodology} 

To evaluate our research questions, we conducted a survey and interview study. All research was conducted after receiving approval from our university's Institutional Review Board.

\subsection{Participant Recruitment}

To recruit a diverse range of participants, we advertised our study on multiple university campuses.
We advertised in student and computing-related Discord servers, Slack channels, and subreddits. We also recruited student organization leaders and faculty to make announcements in their clubs and courses. Participation in the survey was uncompensated and interview participants received a \$10 gift card. All interview participants and survey responses were collected from July-October 2023. 

\subsection{Survey Study}

\subsubsection{Participants} 

We collected 47 survey responses (32 male, 13 female, 2 preferred not to say) from computing students (i.e., students who have taken at least one computing course) across multiple North American universities.
Participants included 2 incoming first-years, 19 first-years, 9 second-years, 12 third-years, and 4 fourth-years. 
42 out of 47 participants had used ChatGPT before but not necessarily other AI tools (e.g. Midjourney, GitHub Copilot). 
Students rated the difficulty of their CS courses from Easy (1) to Hard (5), reporting a median difficulty of 3 ($\mu = 3.0$, $\sigma = 0.78$). 

\subsubsection{Design of survey questions}

To understand students' help-seeking preferences, we asked students to rank seven popular help-seeking resources. These included ChatGPT and Github Copilot which are two popular Generative AI tools used by programmers. These also included five ``traditional'' help-seeking resources based on a previous survey of computing students' help preferences~\cite{wirtz2018resource}. 

Students ranked help received from online resources (e.g., Youtube and StackOverflow), course discussion forums, TAs, instructors, friends, ChatGPT, and GitHub Copilot in order of preference for common CS tasks including understanding course concepts, writing or generating code, debugging, and developing test cases. These scenarios were chosen based on a study that determined what kinds of help computing students seek in TA office hours~\cite{ren2019help} and the stages of programming problem-solving by Loksa et al~\cite{loksa2016programming}. Additionally, participants ranked each resource along the same scale for the following dimensions of assistance: perceived quality, trustworthiness, and timeliness.

Participants were asked to rank their preferences for the 7 resources from ``Most Preferred'' to ``Least Preferred.'' Participants also rated how often they used help-seeking resources on a scale from ``Never'' to ``Hourly.''  

Participants were also asked to rate the help-seeking resources on a five-point Likert-scale for perceived convenience, quality, anonymity, and potential for supporting follow-up questions.
Lastly, participants were asked to evaluate how comfortable they were seeking help from each resource on a five-point Likert-scale. Participants rated their comfort levels with highly anonymous resources (e.g., ChatGPT or online resources) compared to direct resources (e.g., in-class assistance or instructors).

\subsubsection{Survey analysis}
Survey responses were analyzed by computing the mean, median, and standard deviation for all Likert-scale questions and visualizing rankings of the help resources. We present these summary statistics narratively alongside quotes to contextualize the findings. We also generated data visualizations to summarize the ranked preferences. 

\subsection{Interview Study}

\subsubsection{Participants}

To gain a more nuanced perspective on student help-seeking preferences, we also carried out 8 semi-structured interviews (1 female, 7 male) which lasted between 30-45 minutes. All participants had used ChatGPT at least once, but not necessarily other generative AI tools (e.g. Midjourney, GitHub Copilot). 6 out of 8 participants were first or second-year computing students, while the others were third-years or above. Participants were able to schedule an interview without having completed the survey.

\subsubsection{Design of interview questions}

The semi-structured interview format allowed us to deviate from our prepared interview questions, yielding richer and more varied perspectives~\cite{magaldi2020semistructured}. Like the survey, students were asked to rank their help-seeking preferences for each task (e.g. understanding course concepts, writing or generating code, debugging, and developing test cases). We also asked participants to share their help-seeking process, what they valued most and least from each resource, and to draw comparisons with their experiences using ChatGPT. For example, if students claimed they preferred asking ChatGPT due to anonymity, we posed follow-up questions such as: ``If the questions you ask ChatGPT could be viewed by your peers, would you still ask the same questions? Why or why not?'' This allowed us to probe the underlying preferences and values more deeply~\cite{friedman1996value, friedman2013value}.

\subsubsection{Interview analysis}

Interviews were conducted on Zoom, which automatically transcribes the conversation. One researcher parsed through the transcripts to anonymize the participant and make corrections to the transcripts. The transcripts were then analyzed using a thematic analysis technique~\cite{braun2006thematic}. This technique started with open coding~\cite{strauss2004open} with the goal of enabling the emergence of patterns and categories that might not have been anticipated. These codes were grouped thematically. Next, the researchers looked for connections with established theoretical frameworks such as social comparison theory~\cite{suls1991social}, ``recursive try-again loop''~\cite{herring2016academic}, and anonymity effects~\cite{panadero2019empirical, foong2017online}. This approach allowed our qualitative analysis to balance inductive findings grounded in participants experiences and perspectives with existing knowledge about feedback exchange theory. Transcripts were first reviewed individually by three researchers to generate initial codes. Researchers then met twice to collaboratively discuss emerging quotes and themes, refining their collective understanding of the data to achieve convergence and minimize interpretation biases. Researchers came to a consensus through discussion. \td{The results from this thematic analysis are presented in Section~\ref{sec:factors-for-usage}.}

\td{Additionally, we conducted a secondary analysis to support the survey responses. The results for this analysis are presented in Section~\ref{sec:context-of-usage}. For this analysis we adopted a deductive approach that aligned with the questions asked in the survey.} We looked explicitly for instances where participants talked about their usage of help seeking resources including when understanding coding concepts, debugging code, writing code, and creating test cases.

\section{Results}

\begin{table}
\centering
\caption{Frequency of Resource Usage by Students}
\begin{tabular}{lcccccc}
\toprule
Resource & Hourly & Daily & Weekly & Monthly & Never \\
\midrule
Online search & 9 & 24 & 11 & 2 & 1 \\
Friends & 2 & 13 & 16 & 10 & 6 \\
Class forum & 2 & 10 & 14 & 8 & 13 \\
ChatGPT & 4 & 7 & 9 & 11 & 16 \\
Instructor & 1 & 6 & 14 & 17 & 9 \\
TA & 1 & 3 & 15 & 16 & 12 \\
GitHub Copilot & 1 & 1 & 3 & 2 & 40 \\
\bottomrule
\end{tabular}
\label{tab:frequent-usage}
\end{table}

\subsection{Frequency of Usage}
Table~\ref{tab:frequent-usage} displays how frequently students seek help from the seven help-seeking resources in the survey. Searching online was the most popular choice for help seekers with over 70.2\% of students using it on either an hourly or daily basis. Students also relied to a lesser extent on friends (31.9\%), course discussion forums (25.5\%), and ChatGPT (23.4\%) to meet these hourly or daily needs. 
The Internet, friends, and discussion forums were used consistently on a weekly basis by over half of the students surveyed. At least once a week, 44.7\% of students consulted instructors and 42.6\% of students consulted ChatGPT. 

Students depended least on TAs, with 59.6\% of students seeking help monthly or even less frequently. One possible reason for the high frequency use of instructor help could be the inclusion of diversely-sized universities; some universities surveyed had smaller classrooms than others, meaning fewer TAs and more dependence on the instructor. Possible reasons for the reluctance to seek help from TAs could include temporal and spatial limitations, as suggested by previous research~\cite{wirtz2018resource}.  This aligns with what interview participants mentioned regarding resource availability. For example, P2 said, \textit{``For hard classes there's too long of a line; sometimes their [TAs and instructors'] schedules don't fit.''} Multiple students seeking assistance from a single TA or instructor could make this resource seem less accessible and therefore less appealing. 

With regard to the frequency of use, ChatGPT appears to be a slightly polarizing choice; 26.1\% of students who have experience with ChatGPT use it daily while 34.0\% of students never use it at all. ChatGPT's bimodal distribution of usage could be due to a mix of lack of awareness, lack of trust (as further explored later in the interview), or concerns regarding academic dishonesty~\cite{lau2023from, prather2023transformed, zastudil2023generative}.

\begin{figure*}
  \includegraphics[width=0.9\linewidth]{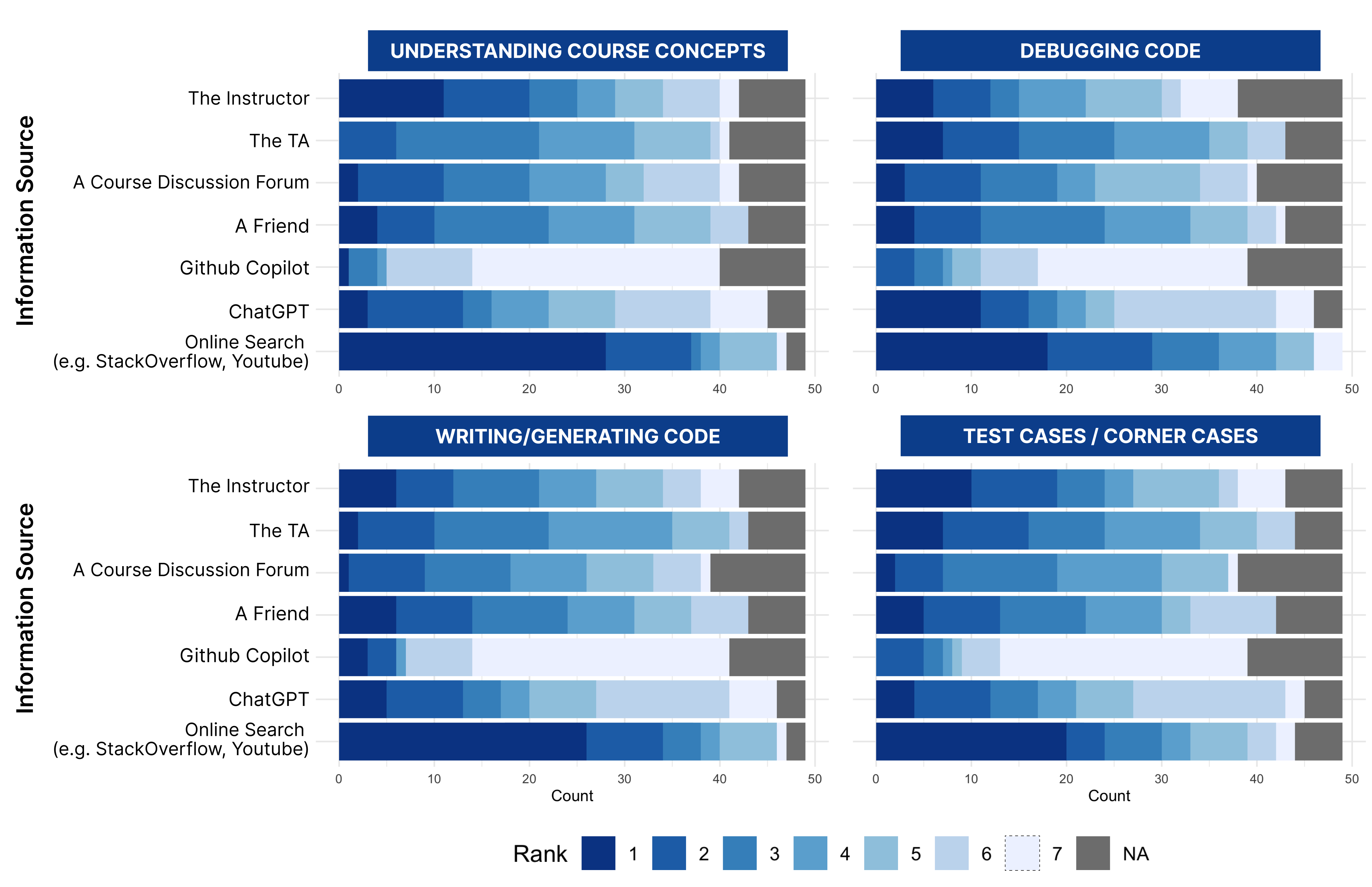}
\caption{Students survey rankings of four common CS tasks}
 \Description{Four stacked barcharts that show the relative ranked preferences of students}
  \label{fig:RQ1}
\end{figure*}


\subsection{Context of Usage}
\label{sec:context-of-usage}

While students may rely more heavily on some help-seeking resources than others, we investigated if specific resources were perceived as more useful for tasks like writing or debugging code. Summarized in Figure~\ref{fig:RQ1}, students' use of various help-seeking resources varied according to task. Across tasks, students ranked the Internet as their most preferred method for seeking help. This is consistent with our observation that students rely heavily on web searches to meet their information needs. The quotes presented in this section came from the interviews and helps to contextualize how generative AI tools were used.

\subsubsection{Learning new concepts}

After internet search, instructors were the next most preferred resources for learning new concepts. However, despite ChatGPT being a relatively nascent resource, it was comparably popular to asking friends or in the discussion forums. The primary reasons for this preference based on our interview data was the iterative nature of responses from ChatGPT and the reduced social pressures of not needing to request help directly from a peer. The following quote demonstrates both aspects:

\begin{quote}
\textit{``It is easier for me to ask ChatGPT because I don’t have to worry... I feel like \textbf{it is kind of rude to continuously ask for more and more details}, especially when you’re approaching someone for help.''} (P1)    
\end{quote}

This quote illustrates how students are at times hesitant to inconvenience their peers. With ChatGPT students can iterate and develop their understanding without concern for social etiquette. Later, the same student shared their biggest pain point when searching online to learn new concepts: 

\begin{quote}
\textit{``The issue is with the scope of what I'm looking for...\textbf{when I try to online search, I can often get sidetracked} by all these complicated, really advanced terms or concepts. So I would try to look at the instructor's items first.''} (P1)
\end{quote}

While internet resources were still highly ranked, P1's comment highlights how for computing students, the web can be an overwhelming source of information, whereas resources like instructors and TAs can provide more scaffolding and guidance. ChatGPT has the potential to bridge this gap by providing more personalized help without the social pressures associated with other help resources. 

P7, who ranked ChatGPT lower than instructors, said that this scaffolding as a `process' was more important than the answer when seeking help: 

\begin{quote}
    \textit{``I feel that [getting an answer instantly] kind of defeats the purpose of me asking a question. If I'm asking a question, I'm trying to understand what's going on. If you just give me the answer, then you’re not describing the process at all. \textbf{I much prefer somebody who gives me the process.} That's what I find most valuable''} (P7).
\end{quote}

These viewpoints emphasize the value of scaffolding help-seeking in learning. How easy it is to navigate and make sense of information can be a reason why students choose certain help-seeking resources over others. 

\subsubsection{Writing code}

When students talked about writing code, a theme emerged that students valued ChatGPT's capacity to support new ideas (P1, P3, P6, P7). One participant pointed out that it felt more challenging to seek abstract, high-level help from a TA: \textit{``I feel like the TA is there for technical help, not for more creative help.''} (P3). Students viewed traditional help resources like the TA as more helpful with common pitfalls and coursework:\textit{``TAs and instructors have experience from multiple students asking questions about the same problems''} (P4). For high-level tasks such as writing code from scratch, P7 noted ChatGPT's creative potential:
\begin{quote}\textit{``If I were to ask ChatGPT a math question, it might get it wrong.\textbf{ But in more creative capacities, that's where I feel like it excels.} When I asked it pre-thinking questions...it'll actually give me a very reasonable list.''}
\end{quote}
While traditional academic scaffolds, such as TAs, remain invaluable for their historical and course-related expertise, participants tended to describe ChatGPT as being more useful when trying to think outside the box.

\subsubsection{Debugging}

Students again strongly preferred internet sources for debugging, followed by ChatGPT and TAs, then course discussion forums. Across the tasks, this appeared to be the use case where ChatGPT was mildy preferred compared to other use cases. Similar to when learning new concepts, students that preferred ChatGPT cited its ability to provide multiple, iterative perspectives and direction in solving specific problems. Another factor was the ability to articulate help requests when debugging. This difficulty of finding relevant solutions when searching online based on an error message was captured by P3: 
\begin{quote}
\textit{``For error messages, when I copy an error message from the terminal and paste it to the Internet, it's not always guaranteed to have a result or a solution.''} 
\end{quote}
In the same breath, the same student highlights an emergent paradigm in debugging scenarios: 
\begin{quote}
\textit{``But if you put it [the error message] into ChatGPT, it will come up with something. \textbf{Even though it may not be right, it will still be helpful, like a correct direction''}} (P3). 
\end{quote}
It is unclear if this participant, \fb{like many students~\cite{guillaume2011mind},} struggled to articulate their error as a search query. It is also possible that models are better able to infer information and omit irrelevant information (e.g.: the working directory or line numbers) from help requests. 

Another participant specifically cited ChatGPT's ability to provide varied responses as helpful:
\begin{quote}
    \textit{``I usually try to give ChatGPT the same prompt a couple times. Based off what it tells me, I'll see what comes up most often. Then, based off of my actual knowledge, I'll see which one's the right answer''} (P1).
\end{quote}

One notable insight is that P1 appears to use frequency as a heuristic for correctness or reliability. If a solution emerges repeatedly from ChatGPT, P1 considered it as more likely to be helpful. However, prompting ChatGPT multiple times with the same prompt may not be the most successful approach. It is possible that---similar to help seeking in other contexts---it is better to reformulate help requests than to repeatedly ask the exact same question. Similar to P3, it is unclear if P1 was able to articulate their queries effectively. Additionally, P1's process of cross-checking ChatGPT's answers with their own knowledge underscores the importance of domain knowledge when evaluating generative AI outputs.

When asked to draw comparisons between internet sources and ChatGPT, one participant observed: \textit{``On Stack Overflow, if somebody asks a question, \textbf{the answer that best suits the question is voted to the top of the list.} And with ChatGPT and Copilot, you don't really get any of that'' (P7).} This again suggests that students require support in evaluating and prioritizing generative AI responses. When seeking help on the web, students have to carefully formulate their search queries to find relevant results. The ability for ChatGPT to infer some meaning from even poorly formed help requests may become an asset when combined with other help seeking resources. 

\begin{figure*}[ht]
  \includegraphics[width=1\textwidth]{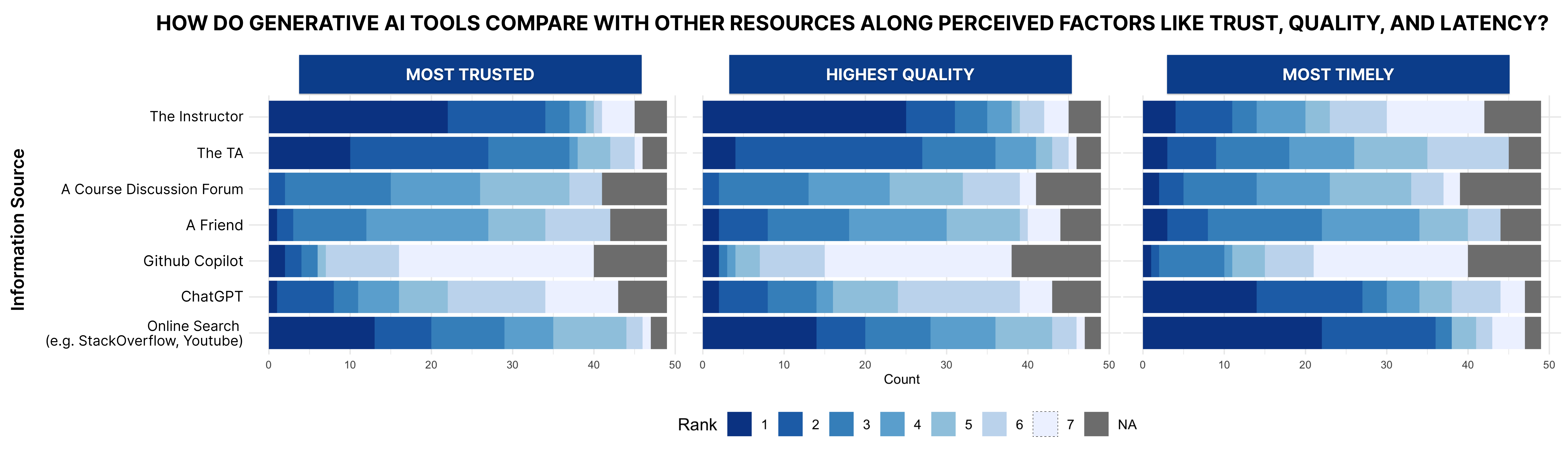}
\caption{Students' survey rankings of perceived qualities of help seeking resources}
 \Description{A visualization of stacked barcharts}
  \label{fig:RQ2}
\end{figure*}

\subsubsection{Developing Test Cases}

\fb{
While not particularly low-ranked in the survey data, participants in the interviews did not talk much about using ChatGPT to find corner cases or to develop test cases. They relied most on course assignment guidelines to test their code and ensure edge cases were covered. P1 said, \textit{``I would look at instructor materials for us to see what kind of test cases... they're specifically asking for.''} This shift in preferences may be due to the computing students sampled for the interview. Students at this level are often overly concerned about the requirements of the assignment, affecting how they face challenging course materials~\cite{paivi2012my}. 
}

\subsection{Factors Influencing Usage}
\label{sec:factors-for-usage}

Based on the ranking data, summarized in Figure~\ref{fig:RQ2}, participants appeared to favor internet resources and ChatGPT over other resources in terms of timeliness; however, these two help resources were perceived as less trustworthy and lower quality than the help received from instructors or TAs. \td{Themes that emerged from interviewed students included trade-offs between rapid feedback and quality, social dynamics, and strategies such as iteration and reformulation when help-seeking.}

\subsubsection{Trust and trade-offs between convenience and quality}

The perceived trade-off between receiving help that is both efficient and accurate emerged as a critical theme from survey and interview data. 68.1\% of the students surveyed Agreed or Strongly Agreed that generative AI tools like ChatGPT were more convenient than alternatives. While they offer immediate and convenient feedback, students noted that this can come at the expense of correctness. 

The lack of feedback latency was considered a major benefit by nearly every interviewee (P1, P3, P4, P5, P6, P7, P8). On the other hand, all but one interview participant described ChatGPT as an unreliable source of help, and a minority of students believed the trade-off between speed and quality was not worth it. First-year students and self-reported novices tended to be harsher in their judgement of ChatGPT's reliability (P3, P4, P6, P8). As P8 candidly shared, \textit{``\textbf{I personally don't really care if [ChatGPT] is fast,} I'll wait [for a peer or TA]...the most important thing is getting it right.''} It is worth noting that P8 also described a negative experience with ChatGPT hallucinations: \textit{``My partner used ChatGPT to generate a Python switch statement, which doesn't exist. I wouldn't trust [ChatGPT] to do it again.''} P2 echoed similar sentiments: \textit{``\textbf{I don't have a lot of trust in ChatGPT} in terms of the harder coding classes and the harder theoretical stuff.''}

Conversely, P7, who had considerable programming experience, found the trade-off between efficiency and quality acceptable: 

\begin{quote}
\textit{``\textbf{I'll take ChatGPT’s advice with a grain of salt anyway.} I will generally use whatever solutions that it gives me...and toss out what doesn’t work.''} 
\end{quote}

\fb{Similarly, P5, a further studies student}, stated: \textit{``I would trust ChatGPT more [than my peers] because you can search it. It has so much data inside. \textbf{[ChatGPT] is like many humans with so much information that you can search through} and find conclusions from.''} 
Another student pointed out that if ChatGPT gave inaccurate information, they would treat it as a \textit{``third party or an alternate input or perspective''} (P4). P4 and P5's insight emphasizes that those who receive the most value from ChatGPT tend to view the model as a search engine or comprehensive resource, from which they must derive pertinent information on their own. This mental model differs from students like P8, who prioritize an expert-level of accuracy.

Although multiple students cited instances when peers gave incorrect or poor quality help, 7 out of 8 students still reportedly preferred seeking help from peers over ChatGPT. This trend aligned with survey data where 51.1\% of students disagreed that ChatGPT provided better quality feedback than alternatives. 
P2 described the conflicting process of seeking help from peers: \textit{``Sometimes they're not even sure about their answers, and when I get a different answer, we're both questioning ourselves.''} P4 went more into detail on this process: \textit{``I'll take the response from my peers if it makes sense to me. I'll take it with a good amount of trust. If I apply it and it doesn't work, that’s OK. The whole purpose is just to get another approach...and if that doesn't work then it’s not a big deal.''} But when he was asked to compare this experience with ChatGPT, he said: \textit{``I definitely don't trust ChatGPT with the accuracy of its response, especially CS stuff.''}

Students largely fell into one of two groups: expecting generative AI tools to be comparable to experts or believing that it is their responsibility to steer the models toward giving useful responses. Unlike when seeking assistance from peers, some participants were less lenient with ChatGPT and subsequently more frustrated when they did not get help after issuing a single prompt. 

\begin{table*}
\centering
\begin{tabular}{lccccc}
\toprule

Scenario & Very Comfortable & Comfortable & Neutral & Uncomfortable & Very Uncomfortable \\
\midrule
Asking ChatGPT for help & 24 (51.1\%) & 8 (17.0\%) & 8 (17.0\%) & 2 (4.3\%) & 5 (10.6\%) \\
Asking a peer privately & 23 (48.9\%) & 11 (23.4\%) & 7 (14.9\%) & 4 (8.5\%) & 2 (4.3\%) \\
Asking a TA/Instructor privately & 19 (40.4\%) & 12 (25.5\%) & 8 (17.0\%) & 8 (17.0\%) & 0 (0\%) \\
Asking other students anonymously & 18 (38.3\%) & 16 (34.0\%) & 4 (8.5\%) & 9 (19.1\%) & 0 (0\%) \\
Asking in front of other students publicly & 12 (25.5\%) & 10 (21.3\%) & 13 (27.7\%) & 9 (19.1\%) & 3 (6.4\%) \\
\bottomrule
\end{tabular}
\caption{Comfort Level of Seeking Help in Different Scenarios}
\label{tab:comfort-levels}
\end{table*}

\subsubsection{Social Aspects}

Social dynamics of help-seeking among computing students involve elements of personal comfort, perceived judgment, and the desire to maintain reciprocal peer relationships. Table~\ref{tab:comfort-levels} indicates that most students reported asking ChatGPT for help as \textit{Very Comfortable}, while other scenarios saw a wider distribution across comfort levels. 68.1\% of respondents reported their comfort levels of asking ChatGPT as \textit{Comfortable} or \textit{Very Comfortable}, compared to traditional forms of help, such as TAs/Instructors (66.0\%) and peers privately (72.3\%) or anonymously (72.3\%). This comfort level dropped to 46.8\% when students were asked how they felt about publicly seeking help. Although students were most comfortable seeking help from their peers, they still found asking ChatGPT more comfortable than asking a TA/Instructor or asking for help in a public setting. Overall, ChatGPT appears to provide a comfortable learning environment for students, free of potential social pressures and apprehensions.

In interviews, one primary concern expressed was the reluctance to `burden' peers with frequent help requests. 

\begin{quote}
    \textit{``I like going to ChatGPT more because I didn't feel like I had to burden my peers at all,''} (P7)
\end{quote}

The desire to avoid straining relationships by over-relying on peers was echoed by most students, such as P1: \textit{``If I ask for help sometimes and then I keep asking 5 more times, I do feel really bad, and I might stop asking.''} P8 admitted that seeking help is challenging and that \textit{``When it comes to people that I know personally, or that I already have messed with, I'll ask...but I still feel judgment.''} Reciprocity considerations were also brought up by participants, most succinctly by P1 who said he does \textit{``try to maintain at least a 50/50-ish balance in terms of asking for help...I don't want our relationship to be this kind of strained—like they're giving a lot more than I'm giving.''} These social dynamics and reputational barriers highlight the continued need for safe student help-seeking spaces. The option to tap into generative AI tools alleviates some fears surrounding being perceived as less knowledgeable or burdensome. The `always-on' nature of AI tools can be empowering, allowing students to progress past common programming sticking points without expending social energy or social capital. P5 aptly puts it: \textit{\textbf{``As an introvert, personally, I think ChatGPT is a very good way to get help} because talking to people kind of takes away my energy.}''

On the other hand, the value of help-seeking in community settings cannot be overstated. While AI tools like ChatGPT could reduce social pressures surrounding help-seeking, students stressed the significant value of human interaction, mutual support, and community. P8 shared an experience that provided testament to the power of collective learning and peer help-seeking, emphasizing the role of mutual struggle in fostering camaraderie: 

\begin{quote}
\textit{``I think \textbf{the thing that got me through CS was the entire community} actively also struggling with their work, learning something new, and having a really, really hard time. I wasn't the only person that was having a hard time...We're all in Discord, we're all in communication...\textbf{I had no problem asking like, `Hey, what did you do?'}'' (P8)}
\end{quote}

\subsubsection{Iteration}

The iterative capabilities of ChatGPT and its ability to handle follow-up questions emerged as a positive for 44.7\% of students, who Agreed or Strongly Agreed that ChatGPT was better at supporting follow-up questions compared to its alternatives. This potential for iteration was especially valuable for students attempting to grasp unfamiliar concepts or clarify misunderstandings. For example, P4 shared the following process: 

\begin{quote}
    \textit{``If you want to ask ChatGPT a concept...I feel like you could ask it something, read its response, and \textbf{ask it to clarify on a response...like follow-up questions}...if I don't understand what it says, I'll prompt it again or I'll ask it to specify.''} 
\end{quote}



Some students valued the autonomy provided by ChatGPT's rapid feedback, enabling them to continuously seek clarifications and review uncertainties at their own pace. The language that students used to describe how they query ChatGPT centers around repetition and continuity, implying that the process of seeking help was not a one-and-done experience but cyclical in nature.

Additionally, students found value in ChatGPT's ability to recall past interactions within a chat session. For example, P8 said:

\begin{quote}
     \textit{``\textbf{ChatGPT has the feature of looking at past conversations} in the thread. I find that very, very helpful. I can make a dedicated thread on one topic.''} 
\end{quote}

This continuity in conversation mirrors the cognitive processes that might take place when students are still conceptualizing new information, allowing them to return to topics and re-familiarize themselves. This constitutes a sensemaking process, reminiscent of organizing notes.

P1 iterated to find answers using the regenerate button:

\begin{quote}
    \textit{``\textbf{I can basically ask ChatGPT as many times as I want.} So that's why I'd often regenerate responses over and over...the answers I found through regenerating responses can be really, really inconsistent.''}
\end{quote}

Instead of reformulating their help request, which is a key strategy in help-seeking~\cite{karabenick2013help}, this student continued asking the same question, expecting an improved answer each time. 

\section{Discussion} 

Our results suggest that while generative AI tools like ChatGPT have become a widely used resource in a very short amount of time, significant barriers still exist that reduce students’ use of it for most tasks. ChatGPT usage varied slightly by task, with students using it more for debugging and writing code. Aligning with prior work, students largely made decisions about which help-seeking resources to use based on the quality and availability of the help resource~\cite{price2017isnap, wirtz2018resource}. However, students highlighted features unique to ChatGPT that traditional resources lacked, such as its creative generative potential and its capacity to  provide iterative support in the form of follow-up questions. Unlike internet resources like Stack Overflow or YouTube, students also appreciated how ChatGPT actively steered them toward solutions for rare bugs or error messages, even if it did not always provide the exact answer. 

Another theme that emerged was related to experience level and trust. Students who were less familiar with these models or who had early negative experiences were much less likely to want to use them. This is partially explained by the concept of calibrated trust~\cite{ashoori2019ai, zhang2020effect} where early negative experiences calibrated students to distrust the models. This is further exacerbated by the fact that models can perform well at times, while also hallucinating incorrect information and struggling on easy multiple choice questions~\cite{savelka2023can, savelka2023large}. Less experienced students described being especially apprehensive about receiving wrong answers and being unable to discern between correct and incorrect responses. This skepticism is a promising finding given the widespread fears about students blindly relying on these tools~\cite{zastudil2023generative, becker2023programming}. 

Conversely, experienced students were more lenient with the models. Students mentioned the necessity of applying their own domain knowledge to evaluate the correctness of the model's responses; hence, more knowledgeable students were better equipped to filter through incorrect responses and find the bits that were valuable or could ``guide’’ their next steps. Across experienced and inexperienced students, distrust did not necessarily mean students failed to receive value from them as we saw most students using the models to varying extents.

Formulating effective help requests appeared to be a common challenge for students. Even without generative AI tools, help-seeking remains a difficult task. When asking peers or instructors for help, students grapple with the 'epistemic paradox of having to know what they do not know' in order to receive assistance~\cite{koole2012the}. This problem appears to persist even with tools like ChatGPT; if students fail to formulate their help request to the model clearly, they do not get the most effective help.  

Given this gap in abilities to use these models, more research is needed to understand how well students are able to use generative AI tools. Instructors need to continue to create pedagogical materials or scaffolding~\cite{jiang2022promptmaker, huang2023memory, macneil2023prompt, denny2023promptly, huang2023causalmapper} that can guide students to maximize the utility of these tools. Crucially, if students are unable to formulate effective help requests, the model will not save them. Models can not infer missing information or pose follow-up questions in the same way instructors might be able to form a theory of mind about what the learner knows and does not know in order to tailor explanations accordingly. Students described instances where they prompted the model and only asked it to regenerate responses when the initial response was not appropriate. On the other hand, confident users described using ChatGPT as a conversation where they needed to reformulate their help requests or provide additional context via follow-up questions.

\subsection{Limitations}

This study offers a important snapshot of how students are initially incorporating generative AI tools into their help-seeking process, providing critical insights into this dynamic landscape. As student usage of these tools is expected to escalate~\cite{becker2023programming, prather2023transformed}, future updates to this study will be essential. Despite the inherent limitations of being one of the first explorations in this rapidly evolving field, our findings shed light on the early stages of student adoption. Notably, our interview study, while relatively smaller (n=8) compared to recent works~\cite{lau2023from, zastudil2023generative} with 18 and 20 participants respectively, is complemented by a medium-sized survey involving 45+ students. However, concentrating on North American students underscores the need for future research with larger and more diverse samples to extend the relevance of our observations to a broader population as the field continues to evolve. \td{Additionally, our sample included students from varied levels of undergraduate study with a skew towards first-year students. Future work could examine the experiences of a more homogeneous population, such as introductory students or even compare the differences between perceptions of introductory computing students versus more senior students and what other help-seeking resources they may rely on.}

\section{Conclusion}

In this study, we examined students' help-seeking behaviors through surveys and interviews. Our findings reveal diverse patterns in the utilization and preferences for generative AI tools in varying tasks. Students described the potential for rapid iteration, creative ideation, and avoiding social pressures as key motivations for using these tools. Notably, the range of preferences and experiences among participants seems closely tied to their individual mental models of these models' function. Students either utilized ChatGPT's responses as an initial step in their problem-solving process or became discouraged after receiving an incorrect answer. Finally, it appears that students received disproportionate benefit when seeking help from generative AI tools. Students who were willing and able to iterate and reformulate their feedback requests described having more success. While these models undoubtedly offer a valuable new resource for help-seeking, it is crucial to acknowledge that the quality of assistance received is intricately tied to the clarity and precision of the initial help request. In this regard, seeking help from generative AI mirrors the importance and difficulty of articulating a help request to a friend~\cite{gao2022you, karabenick2003seeking, newman1990childrens} or formulating and reformulating a search query~\cite{palani2021active, huang2009analyzing, eickhoff2015eye}.

\balance
\bibliographystyle{ACM-Reference-Format}
\bibliography{sample}


\begin{thebibliography}{71}


\ifx \showCODEN    \undefined \def \showCODEN     #1{\unskip}     \fi
\ifx \showDOI      \undefined \def \showDOI       #1{#1}\fi
\ifx \showISBNx    \undefined \def \showISBNx     #1{\unskip}     \fi
\ifx \showISBNxiii \undefined \def \showISBNxiii  #1{\unskip}     \fi
\ifx \showISSN     \undefined \def \showISSN      #1{\unskip}     \fi
\ifx \showLCCN     \undefined \def \showLCCN      #1{\unskip}     \fi
\ifx \shownote     \undefined \def \shownote      #1{#1}          \fi
\ifx \showarticletitle \undefined \def \showarticletitle #1{#1}   \fi
\ifx \showURL      \undefined \def \showURL       {\relax}        \fi
\providecommand\bibfield[2]{#2}
\providecommand\bibinfo[2]{#2}
\providecommand\natexlab[1]{#1}
\providecommand\showeprint[2][]{arXiv:#2}

\bibitem[Aleven et~al\mbox{.}(2003)]%
        {aleven2003help}
\bibfield{author}{\bibinfo{person}{Vincent Aleven}, \bibinfo{person}{Elmar Stahl}, \bibinfo{person}{Silke Schworm}, \bibinfo{person}{Frank Fischer}, {and} \bibinfo{person}{Raven Wallace}.} \bibinfo{year}{2003}\natexlab{}.
\newblock \showarticletitle{Help Seeking and Help Design in Interactive Learning Environments}.
\newblock \bibinfo{journal}{\emph{Review of Educational Research}} \bibinfo{volume}{73}, \bibinfo{number}{3} (\bibinfo{year}{2003}), \bibinfo{pages}{277--320}.
\newblock
\urldef\tempurl%
\url{https://doi.org/10.3102/00346543073003277}
\showDOI{\tempurl}
\showeprint{https://doi.org/10.3102/00346543073003277}


\bibitem[Ashoori and Weisz(2019)]%
        {ashoori2019ai}
\bibfield{author}{\bibinfo{person}{Maryam Ashoori} {and} \bibinfo{person}{Justin~D Weisz}.} \bibinfo{year}{2019}\natexlab{}.
\newblock \showarticletitle{In AI we trust? Factors that influence trustworthiness of AI-infused decision-making processes}.
\newblock \bibinfo{journal}{\emph{arXiv preprint arXiv:1912.02675}} (\bibinfo{year}{2019}).
\newblock


\bibitem[Becker et~al\mbox{.}(2023)]%
        {becker2023programming}
\bibfield{author}{\bibinfo{person}{Brett~A. Becker}, \bibinfo{person}{Paul Denny}, \bibinfo{person}{James Finnie-Ansley}, \bibinfo{person}{Andrew Luxton-Reilly}, \bibinfo{person}{James Prather}, {and} \bibinfo{person}{Eddie~Antonio Santos}.} \bibinfo{year}{2023}\natexlab{}.
\newblock \showarticletitle{Programming Is Hard - Or at Least It Used to Be: Educational Opportunities and Challenges of AI Code Generation}. In \bibinfo{booktitle}{\emph{Proceedings of the 54th ACM Technical Symposium on Computer Science Education V. 1}} (Toronto ON, Canada) \emph{(\bibinfo{series}{SIGCSE 2023})}. \bibinfo{publisher}{Association for Computing Machinery}, \bibinfo{address}{New York, NY, USA}, \bibinfo{pages}{500–506}.
\newblock
\showISBNx{9781450394314}
\urldef\tempurl%
\url{https://doi.org/10.1145/3545945.3569759}
\showDOI{\tempurl}


\bibitem[Bergin et~al\mbox{.}(2005)]%
        {bergin2005examining}
\bibfield{author}{\bibinfo{person}{Susan Bergin}, \bibinfo{person}{Ronan Reilly}, {and} \bibinfo{person}{Desmond Traynor}.} \bibinfo{year}{2005}\natexlab{}.
\newblock \showarticletitle{Examining the Role of Self-Regulated Learning on Introductory Programming Performance}. In \bibinfo{booktitle}{\emph{Proceedings of the First International Workshop on Computing Education Research}} (Seattle, WA, USA) \emph{(\bibinfo{series}{ICER '05})}. \bibinfo{pages}{81–86}.
\newblock
\showISBNx{1595930434}


\bibitem[Braun and Clarke(2006)]%
        {braun2006thematic}
\bibfield{author}{\bibinfo{person}{Virginia Braun} {and} \bibinfo{person}{Victoria Clarke}.} \bibinfo{year}{2006}\natexlab{}.
\newblock \showarticletitle{Using thematic analysis in psychology}.
\newblock \bibinfo{journal}{\emph{Qualitative Research in Psychology}} \bibinfo{volume}{3}, \bibinfo{number}{2} (\bibinfo{year}{2006}), \bibinfo{pages}{77--101}.
\newblock


\bibitem[Buzzetto-Hollywood(2014)]%
        {buzzetto2014youtube}
\bibfield{author}{\bibinfo{person}{Nicole Buzzetto-Hollywood}.} \bibinfo{year}{2014}\natexlab{}.
\newblock \showarticletitle{An Examination of Undergraduate Student’s Perceptions and Predilections of the Use of YouTube in the Teaching and Learning Process}.
\newblock \bibinfo{journal}{\emph{Interdisciplinary Journal of E-Learning and Learning Objects (IJELLO)}}  \bibinfo{volume}{10} (\bibinfo{date}{01} \bibinfo{year}{2014}), \bibinfo{pages}{17--32}.
\newblock
\urldef\tempurl%
\url{https://doi.org/10.28945/1965}
\showDOI{\tempurl}


\bibitem[Cheng and Tsai(2011)]%
        {cheng2011an}
\bibfield{author}{\bibinfo{person}{Kun-Hung Cheng} {and} \bibinfo{person}{Chin-Chung Tsai}.} \bibinfo{year}{2011}\natexlab{}.
\newblock \showarticletitle{An investigation of Taiwan University students' perceptions of online academic help seeking, and their web-based learning self-efficacy}.
\newblock \bibinfo{journal}{\emph{Internet and Higher Education - INTERNET HIGH EDUC}}  \bibinfo{volume}{14} (\bibinfo{date}{07} \bibinfo{year}{2011}), \bibinfo{pages}{150--157}.
\newblock
\urldef\tempurl%
\url{https://doi.org/10.1016/j.iheduc.2011.04.002}
\showDOI{\tempurl}


\bibitem[Chiu et~al\mbox{.}(2006)]%
        {chiu2006understanding}
\bibfield{author}{\bibinfo{person}{Chao-Min Chiu}, \bibinfo{person}{Meng-Hsiang Hsu}, {and} \bibinfo{person}{Eric~T.G. Wang}.} \bibinfo{year}{2006}\natexlab{}.
\newblock \showarticletitle{Understanding knowledge sharing in virtual communities: An integration of social capital and social cognitive theories}.
\newblock \bibinfo{journal}{\emph{Decision Support Systems}} \bibinfo{volume}{42}, \bibinfo{number}{3} (\bibinfo{year}{2006}), \bibinfo{pages}{1872--1888}.
\newblock
\showISSN{0167-9236}


\bibitem[Denny et~al\mbox{.}(2023a)]%
        {denny2023conversing}
\bibfield{author}{\bibinfo{person}{Paul Denny}, \bibinfo{person}{Viraj Kumar}, {and} \bibinfo{person}{Nasser Giacaman}.} \bibinfo{year}{2023}\natexlab{a}.
\newblock \showarticletitle{Conversing with Copilot: Exploring Prompt Engineering for Solving CS1 Problems Using Natural Language}. In \bibinfo{booktitle}{\emph{Proceedings of the 54th ACM Technical Symposium on Computer Science Education V. 1}} (Toronto ON, Canada) \emph{(\bibinfo{series}{SIGCSE 2023})}. \bibinfo{publisher}{Association for Computing Machinery}, \bibinfo{address}{New York, NY, USA}, \bibinfo{pages}{1136–1142}.
\newblock
\showISBNx{9781450394314}
\urldef\tempurl%
\url{https://doi.org/10.1145/3545945.3569823}
\showDOI{\tempurl}


\bibitem[Denny et~al\mbox{.}(2023b)]%
        {denny2023promptly}
\bibfield{author}{\bibinfo{person}{Paul Denny}, \bibinfo{person}{Juho Leinonen}, \bibinfo{person}{James Prather}, \bibinfo{person}{Andrew Luxton-Reilly}, \bibinfo{person}{Thezyrie Amarouche}, \bibinfo{person}{Brett~A Becker}, {and} \bibinfo{person}{Brent~N Reeves}.} \bibinfo{year}{2023}\natexlab{b}.
\newblock \showarticletitle{Promptly: Using prompt problems to teach learners how to effectively utilize ai code generators}.
\newblock \bibinfo{journal}{\emph{arXiv preprint arXiv:2307.16364}} (\bibinfo{year}{2023}).
\newblock


\bibitem[Doebling and Kazerouni(2021)]%
        {doebling2021patterns}
\bibfield{author}{\bibinfo{person}{Augie Doebling} {and} \bibinfo{person}{Ayaan~M. Kazerouni}.} \bibinfo{year}{2021}\natexlab{}.
\newblock \showarticletitle{Patterns of Academic Help-Seeking in Undergraduate Computing Students}. In \bibinfo{booktitle}{\emph{Proceedings of the 21st Koli Calling International Conference on Computing Education Research}} (Joensuu, Finland) \emph{(\bibinfo{series}{Koli Calling '21})}. \bibinfo{publisher}{Association for Computing Machinery}, \bibinfo{address}{New York, NY, USA}, Article \bibinfo{articleno}{13}, \bibinfo{numpages}{10}~pages.
\newblock
\showISBNx{9781450384889}
\urldef\tempurl%
\url{https://doi.org/10.1145/3488042.3488052}
\showDOI{\tempurl}


\bibitem[Dondio and Shaheen(2020)]%
        {dondio2020stackoverflow}
\bibfield{author}{\bibinfo{person}{Pierpaolo Dondio} {and} \bibinfo{person}{Suha Shaheen}.} \bibinfo{year}{2020}\natexlab{}.
\newblock \showarticletitle{Is StackOverflow an Effective Complement to Gaining Practical Knowledge Compared to Traditional Computer Science Learning?}. In \bibinfo{booktitle}{\emph{Proceedings of the 11th International Conference on Education Technology and Computers}} (Amsterdam, Netherlands) \emph{(\bibinfo{series}{ICETC '19})}. \bibinfo{publisher}{ACM}.
\newblock
\showISBNx{9781450372541}


\bibitem[Eickhoff et~al\mbox{.}(2015)]%
        {eickhoff2015eye}
\bibfield{author}{\bibinfo{person}{Carsten Eickhoff}, \bibinfo{person}{Sebastian Dungs}, {and} \bibinfo{person}{Vu Tran}.} \bibinfo{year}{2015}\natexlab{}.
\newblock \showarticletitle{An eye-tracking study of query reformulation}. In \bibinfo{booktitle}{\emph{Proceedings of the 38th international ACM SIGIR conference on research and development in information retrieval}}. \bibinfo{pages}{13--22}.
\newblock


\bibitem[Finnie-Ansley et~al\mbox{.}(2022)]%
        {finnie2022robots}
\bibfield{author}{\bibinfo{person}{James Finnie-Ansley}, \bibinfo{person}{Paul Denny}, \bibinfo{person}{Brett~A Becker}, \bibinfo{person}{Andrew Luxton-Reilly}, {and} \bibinfo{person}{James Prather}.} \bibinfo{year}{2022}\natexlab{}.
\newblock \showarticletitle{The Robots Are Coming: Exploring the Implications of OpenAI Codex on Introductory Programming}. In \bibinfo{booktitle}{\emph{Australasian Computing Education Conf.}} \bibinfo{pages}{10--19}.
\newblock


\bibitem[Fitzgerald et~al\mbox{.}(2008)]%
        {fitzgerald2008debugging}
\bibfield{author}{\bibinfo{person}{Sue Fitzgerald}, \bibinfo{person}{Gary Lewandowski}, \bibinfo{person}{Renee McCauley}, \bibinfo{person}{Laurie Murphy}, \bibinfo{person}{Beth Simon}, \bibinfo{person}{Lynda Thomas}, {and} \bibinfo{person}{Carol Zander}.} \bibinfo{year}{2008}\natexlab{}.
\newblock \showarticletitle{Debugging: finding, fixing and flailing, a multi-institutional study of novice debuggers}.
\newblock \bibinfo{journal}{\emph{Computer Science Education}} \bibinfo{volume}{18}, \bibinfo{number}{2} (\bibinfo{year}{2008}), \bibinfo{pages}{93--116}.
\newblock


\bibitem[Foong et~al\mbox{.}(2017)]%
        {foong2017online}
\bibfield{author}{\bibinfo{person}{Eureka Foong}, \bibinfo{person}{Steven~P Dow}, \bibinfo{person}{Brian~P Bailey}, {and} \bibinfo{person}{Elizabeth~M Gerber}.} \bibinfo{year}{2017}\natexlab{}.
\newblock \showarticletitle{Online feedback exchange: A framework for understanding the socio-psychological factors}. In \bibinfo{booktitle}{\emph{Proceedings of the 2017 CHI Conference on Human Factors in Computing Systems}}. \bibinfo{pages}{4454--4467}.
\newblock


\bibitem[Friedman(1996)]%
        {friedman1996value}
\bibfield{author}{\bibinfo{person}{Batya Friedman}.} \bibinfo{year}{1996}\natexlab{}.
\newblock \showarticletitle{Value-sensitive design}.
\newblock \bibinfo{journal}{\emph{interactions}} \bibinfo{volume}{3}, \bibinfo{number}{6} (\bibinfo{year}{1996}), \bibinfo{pages}{16--23}.
\newblock


\bibitem[Friedman et~al\mbox{.}(2013)]%
        {friedman2013value}
\bibfield{author}{\bibinfo{person}{Batya Friedman}, \bibinfo{person}{Peter~H Kahn}, \bibinfo{person}{Alan Borning}, {and} \bibinfo{person}{Alina Huldtgren}.} \bibinfo{year}{2013}\natexlab{}.
\newblock \showarticletitle{Value sensitive design and information systems}.
\newblock \bibinfo{journal}{\emph{Early engagement and new technologies: Opening up the laboratory}} (\bibinfo{year}{2013}), \bibinfo{pages}{55--95}.
\newblock


\bibitem[Gao et~al\mbox{.}(2022)]%
        {gao2022you}
\bibfield{author}{\bibinfo{person}{Zhikai Gao}, \bibinfo{person}{Bradley Erickson}, \bibinfo{person}{Yiqiao Xu}, \bibinfo{person}{Collin Lynch}, \bibinfo{person}{Sarah Heckman}, {and} \bibinfo{person}{Tiffany Barnes}.} \bibinfo{year}{2022}\natexlab{}.
\newblock \showarticletitle{You Asked, Now What? Modeling Students' Help-Seeking and Coding Actions from Request to Resolution.}
\newblock \bibinfo{journal}{\emph{Journal of Educational Data Mining}} \bibinfo{volume}{14}, \bibinfo{number}{3} (\bibinfo{year}{2022}), \bibinfo{pages}{109--131}.
\newblock


\bibitem[Hellas et~al\mbox{.}(2023)]%
        {hellas2023exploring}
\bibfield{author}{\bibinfo{person}{Arto Hellas}, \bibinfo{person}{Juho Leinonen}, \bibinfo{person}{Sami Sarsa}, \bibinfo{person}{Charles Koutcheme}, \bibinfo{person}{Lilja Kujanpää}, {and} \bibinfo{person}{Juha Sorva}.} \bibinfo{year}{2023}\natexlab{}.
\newblock \showarticletitle{Exploring the Responses of Large Language Models to Beginner Programmers' Help Requests}. In \bibinfo{booktitle}{\emph{Proceedings of the 2023 {ACM} Conference on International Computing Education Research V.1}}. \bibinfo{publisher}{{ACM}}.
\newblock


\bibitem[Herring and Walther(2016)]%
        {herring2016academic}
\bibfield{author}{\bibinfo{person}{Christopher Herring} {and} \bibinfo{person}{Joachim Walther}.} \bibinfo{year}{2016}\natexlab{}.
\newblock \showarticletitle{Academic help-seeking as a stand-alone, metacognitive action: An empirical study of experiences and behaviors in undergraduate engineering students}. In \bibinfo{booktitle}{\emph{2016 ASEE Annual Conference \& Exposition}}.
\newblock


\bibitem[Hou et~al\mbox{.}(2023)]%
        {hou2023more}
\bibfield{author}{\bibinfo{person}{Irene Hou}, \bibinfo{person}{Owen Man}, \bibinfo{person}{Sophie Mettille}, \bibinfo{person}{Sebastian Gutierrez}, \bibinfo{person}{Kenneth Angelikas}, {and} \bibinfo{person}{Stephen MacNeil}.} \bibinfo{year}{2023}\natexlab{}.
\newblock \showarticletitle{More Robots are Coming: Large Multimodal Models (ChatGPT) can Solve Visually Diverse Images of Parsons Problems}.
\newblock \bibinfo{journal}{\emph{arXiv preprint arXiv:2311.04926}} (\bibinfo{year}{2023}).
\newblock


\bibitem[Huang and Efthimiadis(2009)]%
        {huang2009analyzing}
\bibfield{author}{\bibinfo{person}{Jeff Huang} {and} \bibinfo{person}{Efthimis~N Efthimiadis}.} \bibinfo{year}{2009}\natexlab{}.
\newblock \showarticletitle{Analyzing and evaluating query reformulation strategies in web search logs}. In \bibinfo{booktitle}{\emph{Proceedings of the 18th ACM conference on Information and knowledge management}}. \bibinfo{pages}{77--86}.
\newblock


\bibitem[Huang et~al\mbox{.}(2023a)]%
        {huang2023memory}
\bibfield{author}{\bibinfo{person}{Ziheng Huang}, \bibinfo{person}{Sebastian Gutierrez}, \bibinfo{person}{Hemanth Kamana}, {and} \bibinfo{person}{Stephen MacNeil}.} \bibinfo{year}{2023}\natexlab{a}.
\newblock \showarticletitle{Memory sandbox: Transparent and interactive memory management for conversational agents}. In \bibinfo{booktitle}{\emph{Adjunct Proceedings of the 36th Annual ACM Symposium on User Interface Software and Technology}}. \bibinfo{pages}{1--3}.
\newblock


\bibitem[Huang et~al\mbox{.}(2023b)]%
        {huang2023causalmapper}
\bibfield{author}{\bibinfo{person}{Ziheng Huang}, \bibinfo{person}{Kexin Quan}, \bibinfo{person}{Joel Chan}, {and} \bibinfo{person}{Stephen MacNeil}.} \bibinfo{year}{2023}\natexlab{b}.
\newblock \showarticletitle{CausalMapper: Challenging Designers to Think in Systems with Causal Maps and Large Language Model}. In \bibinfo{booktitle}{\emph{Proceedings of the 15th Conference on Creativity and Cognition}} (Virtual Event, USA) \emph{(\bibinfo{series}{C\&C '23})}. \bibinfo{publisher}{Association for Computing Machinery}, \bibinfo{address}{New York, NY, USA}, \bibinfo{pages}{325–329}.
\newblock
\showISBNx{9798400701801}
\urldef\tempurl%
\url{https://doi.org/10.1145/3591196.3596818}
\showDOI{\tempurl}


\bibitem[Jiang et~al\mbox{.}(2022)]%
        {jiang2022promptmaker}
\bibfield{author}{\bibinfo{person}{Ellen Jiang}, \bibinfo{person}{Kristen Olson}, \bibinfo{person}{Edwin Toh}, \bibinfo{person}{Alejandra Molina}, \bibinfo{person}{Aaron Donsbach}, \bibinfo{person}{Michael Terry}, {and} \bibinfo{person}{Carrie~J Cai}.} \bibinfo{year}{2022}\natexlab{}.
\newblock \showarticletitle{PromptMaker: Prompt-based Prototyping with Large Language Models}. In \bibinfo{booktitle}{\emph{CHI Conference on Human Factors in Computing Systems Extended Abstracts}}. \bibinfo{pages}{1--8}.
\newblock


\bibitem[Karabenick(2003)]%
        {karabenick2003seeking}
\bibfield{author}{\bibinfo{person}{Stuart~A Karabenick}.} \bibinfo{year}{2003}\natexlab{}.
\newblock \showarticletitle{Seeking help in large college classes: A person-centered approach}.
\newblock \bibinfo{journal}{\emph{Contemporary educational psychology}} \bibinfo{volume}{28}, \bibinfo{number}{1} (\bibinfo{year}{2003}), \bibinfo{pages}{37--58}.
\newblock


\bibitem[Karabenick and Berger(2013)]%
        {karabenick2013help}
\bibfield{author}{\bibinfo{person}{Stuart~A Karabenick} {and} \bibinfo{person}{Jean-Louis Berger}.} \bibinfo{year}{2013}\natexlab{}.
\newblock \showarticletitle{Help seeking as a self-regulated learning strategy.}
\newblock  (\bibinfo{year}{2013}).
\newblock


\bibitem[Kazemitabaar et~al\mbox{.}(2023)]%
        {kazemitabaar2023studying}
\bibfield{author}{\bibinfo{person}{Majeed Kazemitabaar}, \bibinfo{person}{Justin Chow}, \bibinfo{person}{Carl Ka~To Ma}, \bibinfo{person}{Barbara~J. Ericson}, \bibinfo{person}{David Weintrop}, {and} \bibinfo{person}{Tovi Grossman}.} \bibinfo{year}{2023}\natexlab{}.
\newblock \showarticletitle{Studying the effect of AI Code Generators on Supporting Novice Learners in Introductory Programming}. In \bibinfo{booktitle}{\emph{Proceedings of the 2023 CHI Conference on Human Factors in Computing Systems}} \emph{(\bibinfo{series}{CHI ’23})}. \bibinfo{publisher}{ACM}.
\newblock
\urldef\tempurl%
\url{https://doi.org/10.1145/3544548.3580919}
\showDOI{\tempurl}


\bibitem[Kinnunen and Simon(2012)]%
        {paivi2012my}
\bibfield{author}{\bibinfo{person}{Päivi Kinnunen} {and} \bibinfo{person}{Beth Simon}.} \bibinfo{year}{2012}\natexlab{}.
\newblock \showarticletitle{My program is ok – am I? Computing freshmen's experiences of doing programming assignments}.
\newblock \bibinfo{journal}{\emph{Computer Science Education}} \bibinfo{volume}{22}, \bibinfo{number}{1} (\bibinfo{year}{2012}), \bibinfo{pages}{1--28}.
\newblock
\urldef\tempurl%
\url{https://doi.org/10.1080/08993408.2012.655091}
\showDOI{\tempurl}
\showeprint{https://doi.org/10.1080/08993408.2012.655091}


\bibitem[Koole(2012)]%
        {koole2012the}
\bibfield{author}{\bibinfo{person}{Tom Koole}.} \bibinfo{year}{2012}\natexlab{}.
\newblock \showarticletitle{The epistemics of student problems: Explaining mathematics in a multi-lingual class}.
\newblock \bibinfo{journal}{\emph{Journal of Pragmatics}} \bibinfo{volume}{44}, \bibinfo{number}{13} (\bibinfo{year}{2012}), \bibinfo{pages}{1902--1916}.
\newblock
\showISSN{0378-2166}
\urldef\tempurl%
\url{https://doi.org/10.1016/j.pragma.2012.08.006}
\showDOI{\tempurl}


\bibitem[Kumrow(2007)]%
        {kumrow2007evidence}
\bibfield{author}{\bibinfo{person}{David~E Kumrow}.} \bibinfo{year}{2007}\natexlab{}.
\newblock \showarticletitle{Evidence-based strategies of graduate students to achieve success in a hybrid Web-based course.}
\newblock \bibinfo{journal}{\emph{The Journal of Nursing Education}} \bibinfo{volume}{46}, \bibinfo{number}{3} (\bibinfo{year}{2007}), \bibinfo{pages}{140--145}.
\newblock


\bibitem[Lau and Guo(2023)]%
        {lau2023from}
\bibfield{author}{\bibinfo{person}{Sam Lau} {and} \bibinfo{person}{Philip~J. Guo}.} \bibinfo{year}{2023}\natexlab{}.
\newblock \showarticletitle{From `Ban It Till We Understand It' to "Resistance is Futile": How University Programming Instructors Plan to Adapt as More Students Use AI Code Generation and Explanation Tools such as ChatGPT and GitHub Copilot}. In \bibinfo{booktitle}{\emph{Proceedings of the 2023 ACM Conference on International Computing Education Research V.1 (ICER ’23 V1)}}. \bibinfo{publisher}{ACM}.
\newblock
\urldef\tempurl%
\url{https://doi.org/10.1145/3568813.3600138}
\showDOI{\tempurl}


\bibitem[Leinonen et~al\mbox{.}(2023a)]%
        {leinonen2023comparing}
\bibfield{author}{\bibinfo{person}{Juho Leinonen}, \bibinfo{person}{Paul Denny}, \bibinfo{person}{Stephen MacNeil}, \bibinfo{person}{Sami Sarsa}, \bibinfo{person}{Seth Bernstein}, \bibinfo{person}{Joanne Kim}, \bibinfo{person}{Andrew Tran}, {and} \bibinfo{person}{Arto Hellas}.} \bibinfo{year}{2023}\natexlab{a}.
\newblock \showarticletitle{Comparing Code Explanations Created by Students and Large Language Models}.
\newblock \bibinfo{journal}{\emph{arXiv preprint arXiv:2304.03938}} (\bibinfo{year}{2023}).
\newblock


\bibitem[Leinonen et~al\mbox{.}(2023b)]%
        {leinonen2023using}
\bibfield{author}{\bibinfo{person}{Juho Leinonen}, \bibinfo{person}{Arto Hellas}, \bibinfo{person}{Sami Sarsa}, \bibinfo{person}{Brent Reeves}, \bibinfo{person}{Paul Denny}, \bibinfo{person}{James Prather}, {and} \bibinfo{person}{Brett~A Becker}.} \bibinfo{year}{2023}\natexlab{b}.
\newblock \showarticletitle{Using large language models to enhance programming error messages}. In \bibinfo{booktitle}{\emph{Proceedings of the 54th ACM Technical Symposium on Computer Science Education V. 1}}. \bibinfo{pages}{563--569}.
\newblock


\bibitem[Liffiton et~al\mbox{.}(2023)]%
        {liffiton2023codehelp}
\bibfield{author}{\bibinfo{person}{Mark Liffiton}, \bibinfo{person}{Brad Sheese}, \bibinfo{person}{Jaromir Savelka}, {and} \bibinfo{person}{Paul Denny}.} \bibinfo{year}{2023}\natexlab{}.
\newblock \showarticletitle{Codehelp: Using large language models with guardrails for scalable support in programming classes}.
\newblock \bibinfo{journal}{\emph{arXiv preprint arXiv:2308.06921}} (\bibinfo{year}{2023}).
\newblock


\bibitem[Loksa et~al\mbox{.}(2016)]%
        {loksa2016programming}
\bibfield{author}{\bibinfo{person}{Dastyni Loksa}, \bibinfo{person}{Amy~J Ko}, \bibinfo{person}{Will Jernigan}, \bibinfo{person}{Alannah Oleson}, \bibinfo{person}{Christopher~J Mendez}, {and} \bibinfo{person}{Margaret~M Burnett}.} \bibinfo{year}{2016}\natexlab{}.
\newblock \showarticletitle{Programming, problem solving, and self-awareness: Effects of explicit guidance}. In \bibinfo{booktitle}{\emph{Proceedings of the 2016 CHI conference on human factors in computing systems}}. \bibinfo{pages}{1449--1461}.
\newblock


\bibitem[MacNeil et~al\mbox{.}(2023a)]%
        {macneil2023decoding}
\bibfield{author}{\bibinfo{person}{Stephen MacNeil}, \bibinfo{person}{Paul Denny}, \bibinfo{person}{Andrew Tran}, \bibinfo{person}{Juho Leinonen}, \bibinfo{person}{Seth Bernstein}, \bibinfo{person}{Arto Hellas}, \bibinfo{person}{Sami Sarsa}, {and} \bibinfo{person}{Joanne Kim}.} \bibinfo{year}{2023}\natexlab{a}.
\newblock \showarticletitle{Decoding Logic Errors: A Comparative Study on Bug Detection by Students and Large Language Models}.
\newblock \bibinfo{journal}{\emph{arXiv preprint arXiv:2311.16017}} (\bibinfo{year}{2023}).
\newblock


\bibitem[MacNeil et~al\mbox{.}({[n.\,d.]})]%
        {macneil2022experiences}
\bibfield{author}{\bibinfo{person}{Stephen MacNeil}, \bibinfo{person}{Andrew Tran}, \bibinfo{person}{Arto Hellas}, \bibinfo{person}{Joanne Kim}, \bibinfo{person}{Sami Sarsa}, \bibinfo{person}{Paul Denny}, \bibinfo{person}{Seth Bernstein}, {and} \bibinfo{person}{Juho Leinonen}.} \bibinfo{year}{[n.\,d.]}\natexlab{}.
\newblock \showarticletitle{Experiences from Using Code Explanations Generated by Large Language Models in a Web Software Development E-Book}. In \bibinfo{booktitle}{\emph{Proceedings of the 54th ACM Technical Symposium on Computer Science Education V. 1}} \emph{(\bibinfo{series}{SIGCSE 2023})}. \bibinfo{pages}{931–937}.
\newblock


\bibitem[MacNeil et~al\mbox{.}(2023b)]%
        {macneil2023prompt}
\bibfield{author}{\bibinfo{person}{Stephen MacNeil}, \bibinfo{person}{Andrew Tran}, \bibinfo{person}{Joanne Kim}, \bibinfo{person}{Ziheng Huang}, \bibinfo{person}{Seth Bernstein}, {and} \bibinfo{person}{Dan Mogil}.} \bibinfo{year}{2023}\natexlab{b}.
\newblock \showarticletitle{Prompt Middleware: Mapping Prompts for Large Language Models to UI Affordances}.
\newblock \bibinfo{journal}{\emph{arXiv preprint arXiv:2307.01142}} (\bibinfo{year}{2023}).
\newblock


\bibitem[MacNeil et~al\mbox{.}(2022a)]%
        {macneil2022automatically}
\bibfield{author}{\bibinfo{person}{Stephen MacNeil}, \bibinfo{person}{Andrew Tran}, \bibinfo{person}{Juho Leinonen}, \bibinfo{person}{Paul Denny}, \bibinfo{person}{Joanne Kim}, \bibinfo{person}{Arto Hellas}, \bibinfo{person}{Seth Bernstein}, {and} \bibinfo{person}{Sami Sarsa}.} \bibinfo{year}{2022}\natexlab{a}.
\newblock \showarticletitle{Automatically Generating CS Learning Materials with Large Language Models}.
\newblock \bibinfo{journal}{\emph{arXiv preprint arXiv:2212.05113}} (\bibinfo{year}{2022}).
\newblock


\bibitem[MacNeil et~al\mbox{.}(2022b)]%
        {macneil2022generating}
\bibfield{author}{\bibinfo{person}{Stephen MacNeil}, \bibinfo{person}{Andrew Tran}, \bibinfo{person}{Dan Mogil}, \bibinfo{person}{Seth Bernstein}, \bibinfo{person}{Erin Ross}, {and} \bibinfo{person}{Ziheng Huang}.} \bibinfo{year}{2022}\natexlab{b}.
\newblock \showarticletitle{Generating Diverse Code Explanations Using the GPT-3 Large Language Model}. In \bibinfo{booktitle}{\emph{Proc. of the 2022 ACM Conf. on Int. Computing Education Research - Volume 2}}. \bibinfo{publisher}{ACM}, \bibinfo{pages}{37–39}.
\newblock
\showISBNx{9781450391955}


\bibitem[Magaldi and Berler(2020)]%
        {magaldi2020semistructured}
\bibfield{author}{\bibinfo{person}{Danielle Magaldi} {and} \bibinfo{person}{Matthew Berler}.} \bibinfo{year}{2020}\natexlab{}.
\newblock \showarticletitle{Semi-structured {Interviews}}.
\newblock In \bibinfo{booktitle}{\emph{Encyclopedia of {Personality} and {Individual} {Differences}}}, \bibfield{editor}{\bibinfo{person}{Virgil Zeigler-Hill} {and} \bibinfo{person}{Todd~K. Shackelford}} (Eds.). \bibinfo{publisher}{Springer International Publishing}, \bibinfo{address}{Cham}, \bibinfo{pages}{4825--4830}.
\newblock
\showISBNx{978-3-319-24612-3}
\urldef\tempurl%
\url{https://doi.org/10.1007/978-3-319-24612-3_857}
\showDOI{\tempurl}


\bibitem[Marceau et~al\mbox{.}(2011)]%
        {guillaume2011mind}
\bibfield{author}{\bibinfo{person}{Guillaume Marceau}, \bibinfo{person}{Kathi Fisler}, {and} \bibinfo{person}{Shriram Krishnamurthi}.} \bibinfo{year}{2011}\natexlab{}.
\newblock \showarticletitle{Mind Your Language: On Novices' Interactions with Error Messages}. In \bibinfo{booktitle}{\emph{Proceedings of the 10th SIGPLAN Symposium on New Ideas, New Paradigms, and Reflections on Programming and Software}} (Portland, Oregon, USA) \emph{(\bibinfo{series}{Onward! 2011})}. \bibinfo{pages}{3–18}.
\newblock
\showISBNx{9781450309417}


\bibitem[McCartney et~al\mbox{.}(2016)]%
        {mccartney2016why}
\bibfield{author}{\bibinfo{person}{Robert McCartney}, \bibinfo{person}{Jonas Boustedt}, \bibinfo{person}{Anna Eckerdal}, \bibinfo{person}{Kate Sanders}, \bibinfo{person}{Lynda Thomas}, {and} \bibinfo{person}{Carol Zander}.} \bibinfo{year}{2016}\natexlab{}.
\newblock \showarticletitle{Why Computing Students Learn on Their Own: Motivation for Self-Directed Learning of Computing}.
\newblock \bibinfo{journal}{\emph{ACM Trans. Comput. Educ.}} \bibinfo{volume}{16}, \bibinfo{number}{1}, Article \bibinfo{articleno}{2} (\bibinfo{date}{jan} \bibinfo{year}{2016}), \bibinfo{numpages}{18}~pages.
\newblock
\urldef\tempurl%
\url{https://doi.org/10.1145/2747008}
\showDOI{\tempurl}


\bibitem[Michaeli and Romeike(2019)]%
        {michaeli2019improving}
\bibfield{author}{\bibinfo{person}{Tilman Michaeli} {and} \bibinfo{person}{Ralf Romeike}.} \bibinfo{year}{2019}\natexlab{}.
\newblock \showarticletitle{Improving Debugging Skills in the Classroom: The Effects of Teaching a Systematic Debugging Process}. In \bibinfo{booktitle}{\emph{Proceedings of the 14th Workshop in Primary and Secondary Computing Education}} (Glasgow, Scotland, Uk) \emph{(\bibinfo{series}{WiPSCE '19})}. \bibinfo{publisher}{Association for Computing Machinery}, \bibinfo{address}{New York, NY, USA}, Article \bibinfo{articleno}{15}, \bibinfo{numpages}{7}~pages.
\newblock
\showISBNx{9781450377041}
\urldef\tempurl%
\url{https://doi.org/10.1145/3361721.3361724}
\showDOI{\tempurl}


\bibitem[Nasehi et~al\mbox{.}(2012)]%
        {nasehi2012makes}
\bibfield{author}{\bibinfo{person}{Seyed~Mehdi Nasehi}, \bibinfo{person}{Jonathan Sillito}, \bibinfo{person}{Frank Maurer}, {and} \bibinfo{person}{Chris Burns}.} \bibinfo{year}{2012}\natexlab{}.
\newblock \showarticletitle{What makes a good code example?: A study of programming Q\&A in StackOverflow}. In \bibinfo{booktitle}{\emph{2012 28th IEEE International Conference on Software Maintenance (ICSM)}}. IEEE.
\newblock


\bibitem[Newman(1990)]%
        {newman1990childrens}
\bibfield{author}{\bibinfo{person}{Richard~S. Newman}.} \bibinfo{year}{1990}\natexlab{}.
\newblock \showarticletitle{Children’s help-seeking in the classroom: The role of motivational factors and attitudes.}
\newblock \bibinfo{journal}{\emph{Journal of Educational Psychology}} \bibinfo{volume}{82}, \bibinfo{number}{1} (\bibinfo{year}{1990}), \bibinfo{pages}{71–80}.
\newblock
\urldef\tempurl%
\url{https://doi.org/10.1037/0022-0663.82.1.71}
\showDOI{\tempurl}


\bibitem[Palani et~al\mbox{.}(2021)]%
        {palani2021active}
\bibfield{author}{\bibinfo{person}{Srishti Palani}, \bibinfo{person}{Zijian Ding}, \bibinfo{person}{Stephen MacNeil}, {and} \bibinfo{person}{Steven~P. Dow}.} \bibinfo{year}{2021}\natexlab{}.
\newblock \showarticletitle{The "Active Search" Hypothesis: How Search Strategies Relate to Creative Learning}. In \bibinfo{booktitle}{\emph{Proceedings of the 2021 Conference on Human Information Interaction and Retrieval}} (Canberra ACT, Australia) \emph{(\bibinfo{series}{CHIIR '21})}. \bibinfo{publisher}{Association for Computing Machinery}, \bibinfo{address}{New York, NY, USA}, \bibinfo{pages}{325–329}.
\newblock
\showISBNx{9781450380553}
\urldef\tempurl%
\url{https://doi.org/10.1145/3406522.3446046}
\showDOI{\tempurl}


\bibitem[Panadero and Alqassab(2019)]%
        {panadero2019empirical}
\bibfield{author}{\bibinfo{person}{Ernesto Panadero} {and} \bibinfo{person}{Maryam Alqassab}.} \bibinfo{year}{2019}\natexlab{}.
\newblock \showarticletitle{An empirical review of anonymity effects in peer assessment, peer feedback, peer review, peer evaluation and peer grading}.
\newblock \bibinfo{journal}{\emph{Assessment \& Evaluation in Higher Education}} \bibinfo{volume}{44}, \bibinfo{number}{8} (\bibinfo{year}{2019}), \bibinfo{pages}{1253--1278}.
\newblock


\bibitem[Prather et~al\mbox{.}(2023a)]%
        {prather2023transformed}
\bibfield{author}{\bibinfo{person}{James Prather}, \bibinfo{person}{Paul Denny}, \bibinfo{person}{Juho Leinonen}, \bibinfo{person}{Brett~A Becker}, \bibinfo{person}{Ibrahim Albluwi}, \bibinfo{person}{Michael~E Caspersen}, \bibinfo{person}{Michelle Craig}, \bibinfo{person}{Hieke Keuning}, \bibinfo{person}{Natalie Kiesler}, \bibinfo{person}{Tobias Kohn}, {et~al\mbox{.}}} \bibinfo{year}{2023}\natexlab{a}.
\newblock \showarticletitle{Transformed by Transformers: Navigating the AI Coding Revolution for Computing Education: An ITiCSE Working Group Conducted by Humans}. In \bibinfo{booktitle}{\emph{Proceedings of the 2023 Conference on Innovation and Technology in Computer Science Education V. 2}}. \bibinfo{pages}{561--562}.
\newblock


\bibitem[Prather et~al\mbox{.}(2023b)]%
        {prather2023robots}
\bibfield{author}{\bibinfo{person}{James Prather}, \bibinfo{person}{Paul Denny}, \bibinfo{person}{Juho Leinonen}, \bibinfo{person}{Brett~A Becker}, \bibinfo{person}{Ibrahim Albluwi}, \bibinfo{person}{Michelle Craig}, \bibinfo{person}{Hieke Keuning}, \bibinfo{person}{Natalie Kiesler}, \bibinfo{person}{Tobias Kohn}, \bibinfo{person}{Andrew Luxton-Reilly}, {et~al\mbox{.}}} \bibinfo{year}{2023}\natexlab{b}.
\newblock \showarticletitle{The robots are here: Navigating the generative ai revolution in computing education}.
\newblock \bibinfo{journal}{\emph{arXiv preprint arXiv:2310.00658}} (\bibinfo{year}{2023}).
\newblock


\bibitem[Prather et~al\mbox{.}(2023c)]%
        {prather2023its}
\bibfield{author}{\bibinfo{person}{James Prather}, \bibinfo{person}{Brent~N. Reeves}, \bibinfo{person}{Paul Denny}, \bibinfo{person}{Brett~A. Becker}, \bibinfo{person}{Juho Leinonen}, \bibinfo{person}{Andrew Luxton-Reilly}, \bibinfo{person}{Garrett Powell}, \bibinfo{person}{James Finnie-Ansley}, {and} \bibinfo{person}{Eddie~Antonio Santos}.} \bibinfo{year}{2023}\natexlab{c}.
\newblock \showarticletitle{“It’s Weird That it Knows What I Want”: Usability and Interactions with Copilot for Novice Programmers}.
\newblock \bibinfo{journal}{\emph{ACM Transactions on Computer-Human Interaction}} \bibinfo{volume}{31}, \bibinfo{number}{1} (\bibinfo{date}{Nov.} \bibinfo{year}{2023}), \bibinfo{pages}{1–31}.
\newblock
\showISSN{1557-7325}
\urldef\tempurl%
\url{https://doi.org/10.1145/3617367}
\showDOI{\tempurl}


\bibitem[Price et~al\mbox{.}(2017)]%
        {price2017isnap}
\bibfield{author}{\bibinfo{person}{Thomas~W Price}, \bibinfo{person}{Yihuan Dong}, {and} \bibinfo{person}{Dragan Lipovac}.} \bibinfo{year}{2017}\natexlab{}.
\newblock \showarticletitle{iSnap: towards intelligent tutoring in novice programming environments}. In \bibinfo{booktitle}{\emph{Proc. of the 2017 ACM SIGCSE Technical Symposium on computer science education}}. \bibinfo{pages}{483--488}.
\newblock


\bibitem[Qian and Lehman(2017)]%
        {qian2017students}
\bibfield{author}{\bibinfo{person}{Yizhou Qian} {and} \bibinfo{person}{James Lehman}.} \bibinfo{year}{2017}\natexlab{}.
\newblock \showarticletitle{Students’ Misconceptions and Other Difficulties in Introductory Programming: A Literature Review}.
\newblock \bibinfo{journal}{\emph{ACM Trans. Comput. Educ.}} \bibinfo{volume}{18}, \bibinfo{number}{1}, Article \bibinfo{articleno}{1} (\bibinfo{date}{oct} \bibinfo{year}{2017}), \bibinfo{numpages}{24}~pages.
\newblock
\urldef\tempurl%
\url{https://doi.org/10.1145/3077618}
\showDOI{\tempurl}


\bibitem[Ren et~al\mbox{.}(2019)]%
        {ren2019help}
\bibfield{author}{\bibinfo{person}{Yanyan Ren}, \bibinfo{person}{Shriram Krishnamurthi}, {and} \bibinfo{person}{Kathi Fisler}.} \bibinfo{year}{2019}\natexlab{}.
\newblock \showarticletitle{What Help Do Students Seek in TA Office Hours?}. In \bibinfo{booktitle}{\emph{Proceedings of the 2019 ACM Conference on International Computing Education Research}} (Toronto ON, Canada) \emph{(\bibinfo{series}{ICER '19})}. \bibinfo{publisher}{Association for Computing Machinery}, \bibinfo{address}{New York, NY, USA}, \bibinfo{pages}{41–49}.
\newblock
\showISBNx{9781450361859}
\urldef\tempurl%
\url{https://doi.org/10.1145/3291279.3339418}
\showDOI{\tempurl}


\bibitem[Ryan and Shin(2011)]%
        {ryan2011help}
\bibfield{author}{\bibinfo{person}{Allison~M. Ryan} {and} \bibinfo{person}{Huiyoung Shin}.} \bibinfo{year}{2011}\natexlab{}.
\newblock \showarticletitle{Help-seeking tendencies during early adolescence: An examination of motivational correlates and consequences for achievement}.
\newblock \bibinfo{journal}{\emph{Learning and Instruction}} \bibinfo{volume}{21}, \bibinfo{number}{2} (\bibinfo{year}{2011}), \bibinfo{pages}{247--256}.
\newblock
\showISSN{0959-4752}
\urldef\tempurl%
\url{https://doi.org/10.1016/j.learninstruc.2010.07.003}
\showDOI{\tempurl}
\newblock
\shownote{Special Section I: Solving information-based problems: Evaluating sources and information Special Section II: Stretching the limits in help-seeking research: Theoretical, methodological, and technological advances}.


\bibitem[Sarsa et~al\mbox{.}(2022)]%
        {sarsa2022automatic}
\bibfield{author}{\bibinfo{person}{Sami Sarsa}, \bibinfo{person}{Paul Denny}, \bibinfo{person}{Arto Hellas}, {and} \bibinfo{person}{Juho Leinonen}.} \bibinfo{year}{2022}\natexlab{}.
\newblock \showarticletitle{Automatic Generation of Programming Exercises and Code Explanations Using Large Language Models}. In \bibinfo{booktitle}{\emph{Proc. of the 2022 ACM Conf. on Int. Computing Education Research - Volume 1}}. \bibinfo{publisher}{ACM}, \bibinfo{pages}{27–43}.
\newblock
\showISBNx{9781450391948}


\bibitem[Savelka et~al\mbox{.}(2023a)]%
        {savelka2023thrilled}
\bibfield{author}{\bibinfo{person}{Jaromir Savelka}, \bibinfo{person}{Arav Agarwal}, \bibinfo{person}{Marshall An}, \bibinfo{person}{Chris Bogart}, {and} \bibinfo{person}{Majd Sakr}.} \bibinfo{year}{2023}\natexlab{a}.
\newblock \showarticletitle{Thrilled by Your Progress! Large Language Models (GPT-4) No Longer Struggle to Pass Assessments in Higher Education Programming Courses}.
\newblock  (\bibinfo{year}{2023}), \bibinfo{pages}{78–92}.
\newblock
\showISBNx{9781450399760}
\urldef\tempurl%
\url{https://doi.org/10.1145/3568813.3600142}
\showDOI{\tempurl}


\bibitem[Savelka et~al\mbox{.}(2023b)]%
        {savelka2023large}
\bibfield{author}{\bibinfo{person}{Jaromir Savelka}, \bibinfo{person}{Arav Agarwal}, \bibinfo{person}{Christopher Bogart}, {and} \bibinfo{person}{Majd Sakr}.} \bibinfo{year}{2023}\natexlab{b}.
\newblock \bibinfo{title}{Large Language Models (GPT) Struggle to Answer Multiple-Choice Questions about Code}.
\newblock
\newblock
\showeprint[arxiv]{2303.08033}~[cs.CL]


\bibitem[Savelka et~al\mbox{.}(2023c)]%
        {savelka2023can}
\bibfield{author}{\bibinfo{person}{Jaromir Savelka}, \bibinfo{person}{Arav Agarwal}, \bibinfo{person}{Christopher Bogart}, \bibinfo{person}{Yifan Song}, {and} \bibinfo{person}{Majd Sakr}.} \bibinfo{year}{2023}\natexlab{c}.
\newblock \showarticletitle{Can Generative Pre-trained Transformers (GPT) Pass Assessments in Higher Education Programming Courses?}
\newblock \bibinfo{journal}{\emph{arXiv preprint arXiv:2303.09325}} (\bibinfo{year}{2023}).
\newblock


\bibitem[Sharif and Zainab(2004)]%
        {sharif2004undergraduates}
\bibfield{author}{\bibinfo{person}{Mohd Sharif} {and} \bibinfo{person}{AN Zainab}.} \bibinfo{year}{2004}\natexlab{}.
\newblock \showarticletitle{Undergraduates in computer science and information technology using the Internet as a Resource}.
\newblock \bibinfo{journal}{\emph{Malaysian Journal of Library \& Information Science}} \bibinfo{volume}{9}, \bibinfo{number}{1} (\bibinfo{year}{2004}), \bibinfo{pages}{1--16}.
\newblock


\bibitem[Strauss and Corbin(2004)]%
        {strauss2004open}
\bibfield{author}{\bibinfo{person}{Anselm~L Strauss} {and} \bibinfo{person}{Juliet Corbin}.} \bibinfo{year}{2004}\natexlab{}.
\newblock \showarticletitle{Open coding}.
\newblock \bibinfo{journal}{\emph{Social research methods: A reader}} (\bibinfo{year}{2004}), \bibinfo{pages}{303--306}.
\newblock


\bibitem[Suls and Wills(1991)]%
        {suls1991social}
\bibfield{author}{\bibinfo{person}{Jerry~Ed Suls} {and} \bibinfo{person}{Thomas Ashby~Ed Wills}.} \bibinfo{year}{1991}\natexlab{}.
\newblock \bibinfo{booktitle}{\emph{Social comparison: Contemporary theory and research.}}
\newblock \bibinfo{publisher}{Lawrence Erlbaum Associates, Inc}.
\newblock


\bibitem[Tran et~al\mbox{.}(2023)]%
        {tran2023using}
\bibfield{author}{\bibinfo{person}{Andrew Tran}, \bibinfo{person}{Linxuan Li}, \bibinfo{person}{Egi Rama}, \bibinfo{person}{Kenneth Angelikas}, {and} \bibinfo{person}{Stephen MacNeil}.} \bibinfo{year}{2023}\natexlab{}.
\newblock \showarticletitle{Using Large Language Models to Automatically Identify Programming Concepts in Code Snippets}. In \bibinfo{booktitle}{\emph{Proc. of the 2023 ACM Conf. on Int. Computing Education Research - Volume 2}}, Vol.~\bibinfo{volume}{1}. \bibinfo{publisher}{ACM}, \bibinfo{pages}{563--569}.
\newblock
\urldef\tempurl%
\url{https://doi.org/10.1145/3568812.3603482}
\showDOI{\tempurl}


\bibitem[Vasilescu et~al\mbox{.}(2013)]%
        {vasilescu2013stackoverflow}
\bibfield{author}{\bibinfo{person}{Bogdan Vasilescu}, \bibinfo{person}{Vladimir Filkov}, {and} \bibinfo{person}{Alexander Serebrenik}.} \bibinfo{year}{2013}\natexlab{}.
\newblock \showarticletitle{Stackoverflow and github: Associations between software development and crowdsourced knowledge}. In \bibinfo{booktitle}{\emph{2013 International Conference on Social Computing}}. IEEE, \bibinfo{pages}{188--195}.
\newblock


\bibitem[Wirtz et~al\mbox{.}(2018)]%
        {wirtz2018resource}
\bibfield{author}{\bibinfo{person}{Elizabeth Wirtz}, \bibinfo{person}{Amy Dunford}, \bibinfo{person}{Edward Berger}, \bibinfo{person}{Elizabeth Briody}, \bibinfo{person}{Gireesh Guruprasad}, {and} \bibinfo{person}{Ryan Senkpeil}.} \bibinfo{year}{2018}\natexlab{}.
\newblock \showarticletitle{Resource usage and usefulness: academic help-seeking behaviours of undergraduate engineering students}.
\newblock \bibinfo{journal}{\emph{Australasian Journal of Engineering Education}} \bibinfo{volume}{23}, \bibinfo{number}{2} (\bibinfo{year}{2018}), \bibinfo{pages}{62--70}.
\newblock
\urldef\tempurl%
\url{https://doi.org/10.1080/22054952.2018.1525889}
\showDOI{\tempurl}
\showeprint{https://doi.org/10.1080/22054952.2018.1525889}


\bibitem[Yin et~al\mbox{.}(2021)]%
        {yin2021conversation}
\bibfield{author}{\bibinfo{person}{Jiaqi Yin}, \bibinfo{person}{Tiong-Thye Goh}, \bibinfo{person}{Bing Yang}, {and} \bibinfo{person}{Yang Xiaobin}.} \bibinfo{year}{2021}\natexlab{}.
\newblock \showarticletitle{Conversation Technology With Micro-Learning: The Impact of Chatbot-Based Learning on Students’ Learning Motivation and Performance}.
\newblock \bibinfo{journal}{\emph{Journal of Educational Computing Research}} \bibinfo{volume}{59}, \bibinfo{number}{1} (\bibinfo{year}{2021}), \bibinfo{pages}{154--177}.
\newblock
\urldef\tempurl%
\url{https://doi.org/10.1177/0735633120952067}
\showDOI{\tempurl}


\bibitem[Zander et~al\mbox{.}(2012)]%
        {zander2012self}
\bibfield{author}{\bibinfo{person}{Carol Zander}, \bibinfo{person}{Jonas Boustedt}, \bibinfo{person}{Anna Eckerdal}, \bibinfo{person}{Robert McCartney}, \bibinfo{person}{Kate Sanders}, \bibinfo{person}{Jan~Erik Mostr\"{o}m}, {and} \bibinfo{person}{Lynda Thomas}.} \bibinfo{year}{2012}\natexlab{}.
\newblock \showarticletitle{Self-Directed Learning: Stories from Industry}. In \bibinfo{booktitle}{\emph{Proceedings of the 12th Koli Calling International Conference on Computing Education Research}} (Koli, Finland) \emph{(\bibinfo{series}{Koli Calling '12})}. \bibinfo{publisher}{Association for Computing Machinery}, \bibinfo{address}{New York, NY, USA}, \bibinfo{pages}{111–117}.
\newblock
\showISBNx{9781450317955}
\urldef\tempurl%
\url{https://doi.org/10.1145/2401796.2401810}
\showDOI{\tempurl}


\bibitem[Zastudil et~al\mbox{.}(2023)]%
        {zastudil2023generative}
\bibfield{author}{\bibinfo{person}{Cynthia Zastudil}, \bibinfo{person}{Magdalena Rogalska}, \bibinfo{person}{Christine Kapp}, \bibinfo{person}{Jennifer Vaughn}, {and} \bibinfo{person}{Stephen MacNeil}.} \bibinfo{year}{2023}\natexlab{}.
\newblock \showarticletitle{Generative AI in Computing Education: Perspectives of Students and Instructors}.
\newblock \bibinfo{journal}{\emph{arXiv preprint arXiv:2308.04309}} (\bibinfo{year}{2023}).
\newblock


\bibitem[Zhang et~al\mbox{.}(2020)]%
        {zhang2020effect}
\bibfield{author}{\bibinfo{person}{Yunfeng Zhang}, \bibinfo{person}{Q~Vera Liao}, {and} \bibinfo{person}{Rachel~KE Bellamy}.} \bibinfo{year}{2020}\natexlab{}.
\newblock \showarticletitle{Effect of confidence and explanation on accuracy and trust calibration in AI-assisted decision making}. In \bibinfo{booktitle}{\emph{Proceedings of the 2020 conference on fairness, accountability, and transparency}}. \bibinfo{pages}{295--305}.
\newblock


\end{thebibliography}


\end{document}